\documentclass[twocolumn,showpacs,preprintnumbers,amsmath,amssymb,prb,superscriptaddress]{revtex4-1}
\usepackage{graphicx}
\usepackage{dcolumn}
\usepackage{bm}
\usepackage{epstopdf}
\usepackage{xcolor}
\usepackage{appendix}

\usepackage[colorlinks=true,linkcolor=blue, citecolor=blue]{hyperref}%

\begin{document}

\title{Magnetic impurities on superconducting Pb surfaces}

\author{Ming-Hung Wu}%
\affiliation{University of Bristol, School of Physics, HH Wills Physics Laboratory, Tyndall Avenue, Bristol BS8 1TL, England}
\author{Emma Thill}
\affiliation{University of Bristol, School of Physics, HH Wills Physics Laboratory, Tyndall Avenue, Bristol BS8 1TL, England}
\author{Jacob Crosbie}
\affiliation{University of Bristol, School of Physics, HH Wills Physics Laboratory, Tyndall Avenue, Bristol BS8 1TL, England}
\author{Tom G. Saunderson}
\affiliation{Johannes Gutenberg Universit\"at Mainz, Institut f\"ur Physik, Staudingerweg 7, 55128 Mainz, Germany}
\affiliation{University of Bristol, School of Physics, HH Wills Physics Laboratory, Tyndall Avenue, Bristol BS8 1TL, England} 
\author{Martin Gradhand}
\affiliation{Johannes Gutenberg Universit\"at Mainz, Institut f\"ur Physik, Staudingerweg 7, 55128 Mainz, Germany}
\affiliation{University of Bristol, School of Physics, HH Wills Physics Laboratory, Tyndall Avenue, Bristol BS8 1TL, England}


\date{\today}

\begin{abstract}
It has been predicted theoretically and found experimentally that magnetic impurities induce localized bound states within the superconducting energy gap, called Yu-Shiba-Rusinov (YSR) states. Combining symmetry analysis with experimental findings provides a convincing argument for the energy splitting and distribution of the YSR peaks, but the full details of the electronic structure remain elusive and simple models with point scatterers lack the full orbital complexity required to meet this challenge. In this work we combine a Green's function based first-principles method, which incorporates a phenomenological parameterization of the superconducting state, with orbitally complex impurity potentials to make material-specific predictions of realistic systems.  We study the effect of 3$d$ transition elements on the superconducting energy gap of a Pb (001) surface. Not only do we find a good agreement with experiment, we also show that the energetic position, strength and orbital composition of the YSR states depend strongly on the chemical makeup of the impurity and its position with respect to the surface. Such quantitative results cannot be derived from simplified models but require full material specific calculations.
\end{abstract}

\pacs{Valid PACS appear here}
\maketitle


\section{\label{sec:level1} Introduction}

The effect of impurities on the superconducting energy gap has attracted a lot of interest, both experimentally \cite{Ronen_2016,Hoffman_2003,Hirschfeld_2015,Avraham_2018,Hoffman_2002,Ruby2016,Schmid2022,Beck2021,Liebhaber2020} as well as theoretically \cite{Steiner2022,Qiang_2003,Sulangi_2018,Choi_2017}. In particular, magnetic impurities induce in-gap states, also known as Yu-Shiba-Rusinov (YSR) bound states, described as localized excitations within the gap, who's energetic positioning and strength change depending on the chemical composition of both the superconductor and the impurity \cite{Zhen_2012, Fowler_1970, Hanaguri_2010, balatsky_2006, Allan_2012}. These states have received special attention from the field of quantum computing, as a coupled chain of magnetic impurities on the surface of an $s$-wave superconductor could give rise to zero mode Majorana fermions \cite{Choy2011,Ruby_2015_endstates,Beenakker_2013,Alicea_2012, Elliot_2015,Schneider2022}, enabling the construction of noise resilient qubits \cite{Pawlak_2019,Choi_2019} which are potential building blocks of topological quantum computers \cite{Ruby_2015_endstates,Beenakker_2013,Alicea_2012, Elliot_2015, Pawlak_2019, Choi_2019}. In order to achieve this ambition such states need to be carefully optimized requiring a detailed understanding of the microscopic mechanism.

YSR states form due to the exchange interaction between the impurity's spin and the Cooper pairs in the superconductor \cite{Abriskov1960,Phillips1963,Yu_1965,shiba_1968,Rusinov_1969_superconductivity,HEINRICH_2018}, causing them to break apart. This results in a Bogoliubov quasiparticle, a superposition of the excited electron and the hole left behind. The quasiparticles will scatter off the impurity potential, interfering with themselves, leading to distinct quasiparticle interference (QPI) patterns. Measured via scanning tunnelling microscopy (STM), such real space QPI patterns give access to the underlying local density of states (LDOS), including in-gap features. \cite{Ronen_2016,Hirschfeld_2015,Avraham_2018,Fischer_2007,Ji_2008,Ruby2016,Hoffman_2002}  

Theoretically, this interference pattern is often described modelling the impurity as a point scatterer with an isotropic potential \cite{Hirschfeld2011,Hirschfeld_2015,Kreisel2017,pereg_2008,Sulangi2017,VanDyke2016,Qiang_2003,Wang_2003,Zhang_2009,Choi_2017,Sulangi_2018}, entirely working in a band picture in reciprocal space. In order to make contact to experiment the QPI data is Fourier transformed\cite{Hoffman_2003} and compared to the theoretical models.

We report an implementation of the superconducting density functional theory \cite{Olivera_1988} in a self-consistent Korringa-Kohn-Rostoker (KKR) Green's function method \cite{Korringa_1947,Kohn_1954,csire_2017,Saunderson_2020} working in real space incorporating the full orbital character of substitutional impurities \cite{Saunderson2020b,Saunderson_2022}. This allows full access to the real space modulations on the surface of a superconductor. As we are able to consider chemically different impurities and their orbital character we can resolve element-specific changes in the QPI. This allows for a more direct comparison to STM experiments and a deeper material specific understanding of the impurity level formation in superconducting materials.

After a brief introduction of the computational method as well as the known principle behind the formation of in-gap states induced by magnetic impurities, we start by discussing results of a single Cobalt impurity as an adatom as well as an embedded surface atom on a superconducting Pb surface. This will serve as a prelude to the crystal field splitting of the YSR bound states relevant throughout this work. We will extend this discussion to all different 3$d$ transition elements, ranging from Scandium ($Z=21$) to Zinc ($Z=39$) in order to draw general conclusions and understand the trends in formation of YSR states.

\section{\label{sec:CM} Computational Method}

\begin{figure*}[ht!]
    \centering
    \includegraphics[width=1\linewidth,clip]{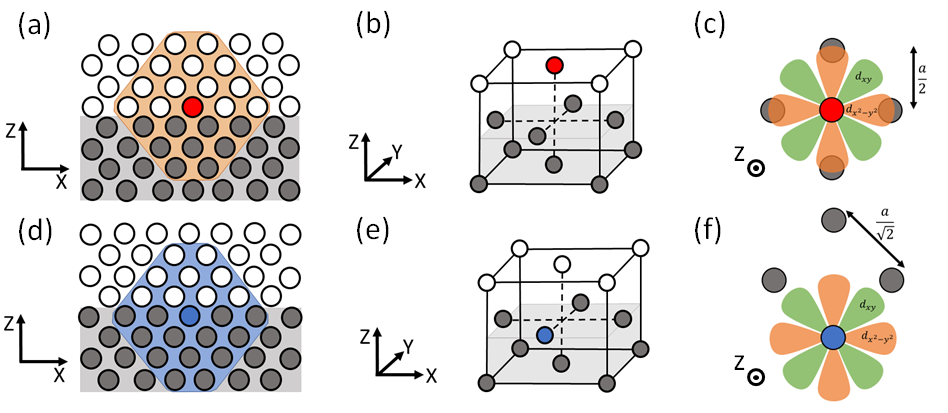}
    \caption{Schematics of the geometric and structural configurations of the semi-infinite Pb thin-film, with Pb atoms shown in grey and vacuum sites in white. (a) The red disk indicates the adatom position with the orange area representing the real space impurity cluster. (b) Adatom impurity (red) shown with respect to the face-centered cubic (fcc) unit cell. (c) Orientation of the $d_{xy}$ (green) and $d_{x^2-y^2}$ (orange) orbitals with respect to the surface layer Pb atoms below for the adatom (d) The equivalent to (a), but for the surface impurity (blue) and the impurity cluster highlighted in light blue. (e) The surface impurity position (blue) with respect to the fcc unit cell. (f) Orientation of the $d_{xy}$ (green) and $d_{x^2-y^2}$ (orange) orbitals with respect to the surface layer Pb atoms below for the surface atom.}
    \label{fig:1}
\end{figure*}

In order to access the real space modulations in the LDOS we perform self-consistent superconducting density functional theory \cite{Olivera_1988} calculations on an impurity cluster embedded in the perfect periodic surface of a Pb crystal. For the solution of the scalar relativistic Bogoliubov-de Gennes (BdG) equations from first principles we use a Korringa-Kohn-Rostoker (KKR) Green's function method \cite{csire_2017,Saunderson_2020,Saunderson2020b,Saunderson_2022}. The relevant accuracy around the Fermi energy requires a semi-circle contour in the upper plane of the complex energy of typically $50$ energy points. Furthermore, the localized basis set gives direct access to an orbital resolved \textcolor{black}{local density of states (LPDOS)} of the bound states inside the superconducting gap. \textcolor{black}{This means that we are calculating both the local and partial DOS, which we will refer to as the LPDOS.} Beyond the description of impurities \cite{Saunderson2020b,Saunderson_2022,Nyari2021}, this methodology has already been shown to successfully model superconductor-metal interfaces \cite{Csire2018}, superconductor-topological insulator interfaces \cite{Park2020a,Russmann2022a} and unconventional pairing \cite{Csire2018a,Ghosh2019a}.

The surfaces are modelled by $9$ layers of Lead sandwiched between \textcolor{black}{4 layers of vacuum on each side adjacent to semi infinite vacuum leads considered by curtailing any couplings beyond the considered 17 layers}. The Pb crystal structure is fcc in (001) direction with a lattice constant of $4.96$~\AA. For the single impurity calculations we consider a real space cluster of $87$ Pb atoms and vacuum sites centred around the substitutional impurity atom which is either an adatom or a surface atom as depicted in Fig.~\ref{fig:1}(a) and (d) respectively. The adatom is placed at a height $h=a/2=2.47$~\AA \ above the Pb surface occupying a vacuum site (see Fig.~\ref{fig:1}(a) and (b)), whereas the surface impurity is substituted for a Pb surface atom as shown in Fig.~\ref{fig:1}(d) and (e). 

Within our method, \textcolor{black}{which so far has been used to model the bulk,}\cite{Saunderson_2020} the superconducting interaction parameter is free and typically tuned to fit the experimentally observed gap sizes. Here, \textcolor{black}{when modelling the surface,} we set it to $\Lambda = 4.73$~eV recovering gap sizes comparable to Ruby \textit{et al.}~[\onlinecite{Ruby_2015_lead2band}], \textcolor{black}{and set the interaction parameter to zero on the vacuum sites}. For all LPDOS calculations, which even for the impurity cluster require an initial calculation of the Green's function of the periodic crystal, we used an imaginary part of the energy of $1.35 \cdot 10^{-5}$~eV with an average of $200\times200$ $k$-points in the 2-dimensional Brillouin zone. Such a dense mesh was required to resolve the fine structure of the LPDOS inside the superconducting gap on the sub meV energy scale.

\begin{figure*}[ht!]
\includegraphics[width=16cm]{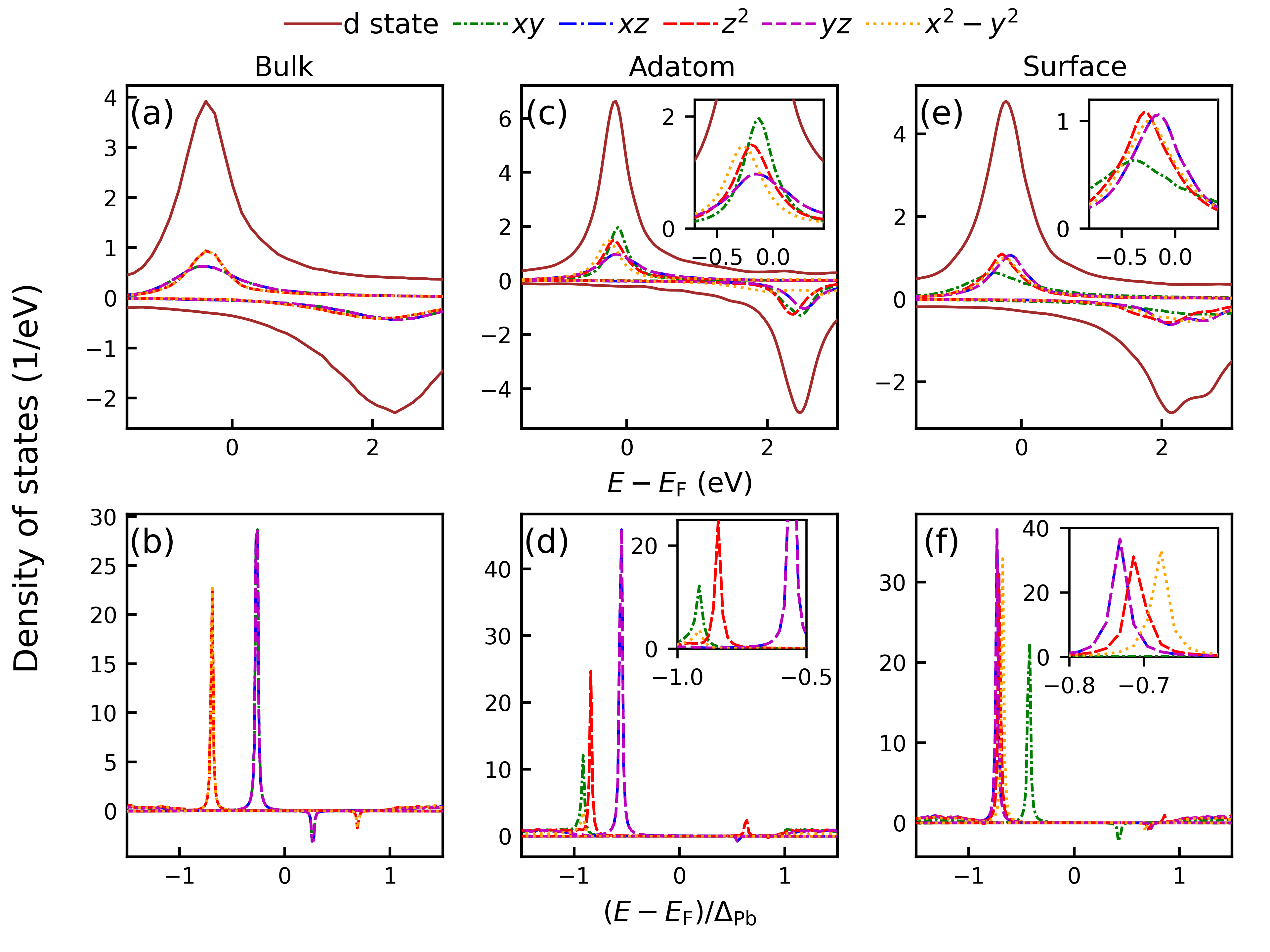}
\caption{Normal (a, c, e) and superconducting state (b, d, f) $d$ orbital resolved LDOS of Vanadium as an impurity in bulk Pb (a and b), as an adatom (c and d) and a surface atom (e and f). The $d_{xz}/d_{yz}$ orbitals are perfectly degenerate. The spin-up LPDOS is shown in the positive part of the $y$-axis, while the spin-down LPDOS is shown in the negative part of the $y$-axis.}
\label{fig:2}
\end{figure*}

\section{\label{sec:normal_state} YSR States and the Crystal Field}

The principle formation of YSR in-gap states induced by magnetic impurities in superconducting materials is well understood theoretically.~\cite{Schrieffer_1967, Yu_1965, shiba_1968, Rusinov_1969_superconductivity, HEINRICH_2018, Matsuura_1977, Flatte1997,Fetter1965} They have been described theoretically using simplified models~\cite{Yu_1965, shiba_1968, Rusinov_1969_superconductivity}, they have been identified experimentally in various scenarios~\cite{Franke_2011, Yazdani_1997, Binning_1987, Binning_1983, Ruby2016, Ruby_2015_endstates, Ruby_2018_wavefunction,Cornils_2017,Yazdani_1997} and their explicit energetic positions have been found within material specific calculations.~\cite{Saunderson_2022, Choi_2017, Kamlapure_2021, Beenakker_2013} It has been shown that their exact energetic positions depend quite dramatically on detail of the electronic structure and the hybridization between the impurity and host crystal electronic states.~\cite{Saunderson_2022,von_2021} For that reason any discussion has often been limited to symmetry derived arguments accounting for the principle splitting of impurity states.~\cite{Saunderson_2022} In the case of 3$d$ impurities, typically required for inducing magnetic moments, this implies a splitting into $e_g$ and $t_{2g}$ states as determined by the crystal field within cubic crystals.~\cite{Saunderson_2022,Schrieffer_1967} However, as many of the experimental investigations of YSR in-gap states will rely on probing surfaces with its further reduced symmetries, the crystal field splitting will lead to further splitting of the in-gap states.~\cite{Ruby2016,Ruby_2015_endstates,Franke_2011} In such a situation only the $d_{xz}$ and $d_{yz}$ might remain degenerate.~\cite{Ruby2016} In that context it is important to realize that also the in-plane position of any impurity will fundamentally change the energetic ordering of the crystal-field split states.

In Fig.~\ref{fig:1}(c) and (f) we show the comparison between the hollow site position of an adatom (c) as well as the substitutional position of an in-plane surface impurity (g). Not only does the nearest neighbour distance between the impurity and the first host atom changes considerably but also the $d_{xy}$ and $d_{x^2-y^2}$ orbital will change their respective meaning. While for the hollow site position the lobes of the $d_{x^2-y^2}$ will point along the nearest neighbour bonds it will be the $d_{xy}$ orbitals pointing along the bonds for the substitutional impurity position. Such a situation will change the hybridization between the crystal electronic states as well as the impurity levels leading to a change in the effective crystal field and subsequently in changed energetic arrangement for the YSR in-gap states.

\section{\label{sec:citeref} Results}

\subsection{\label{sec:normal_state} Vanadium at the surface of superconducting Pb}

\begin{figure*}[ht!]
    \centering
    \hspace{-1cm}\includegraphics[width = 
    \linewidth]{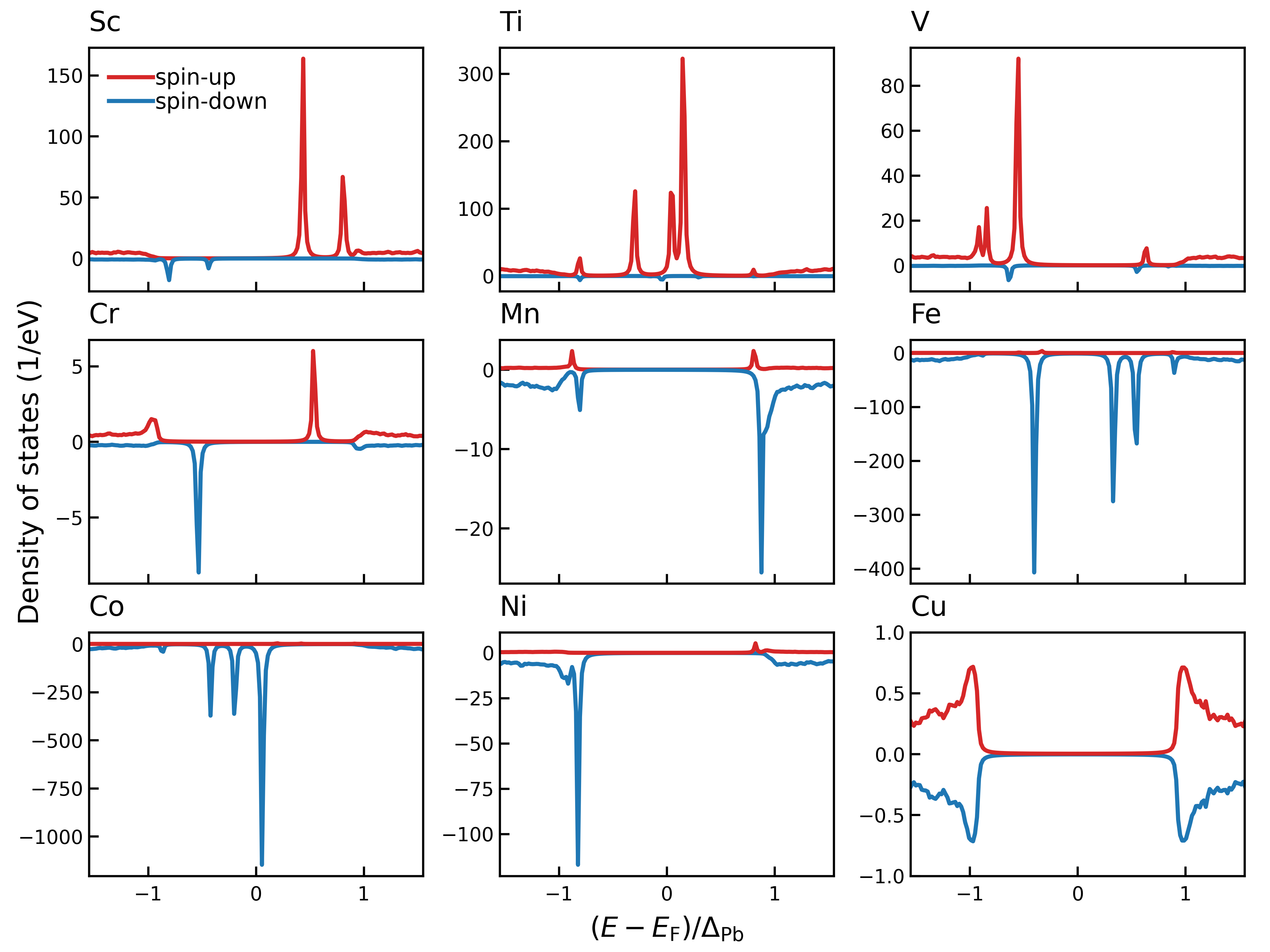}
    \caption{Superconducting state total \textcolor{black}{spin-resolved} LDOS of the 3$d$ transition elements Scandium (21) through to Copper (29) as single adatom impurities to a Pb (001) surface. \textcolor{black}{ The spin-up LDOS is shown in the positive part of the y-axis, while the spin-down LDOS is shown in the negative part of the y-axis.} Each figure has the same $x$ axis.}
    \label{fig:3}
\end{figure*}

In a first application we investigate a Vanadium impurity on the surface of superconducting Pb, where the results are summarized in Fig.~\ref{fig:2} for the bulk system ((a) and (b)), the adatom configuration ((c) and (d)), as well as the surface impurity ((e) and (f)). As predicted we observe a four-fold splitting of the degeneracy of the $d$ orbitals in the case of both the adatom (Fig.~\ref{fig:2}(c) and (d)) and the surface atom (Fig.~\ref{fig:2}(e) and (f)): (1) $d_{z^2}$, (2) $d_{x^2-y^2}$, (3) $d_{xz}/d_{yz}$ and (4) $d_{xy}$ as soon as the surface breaks the conventional cubic symmetry of the 3$d$ crystal. In contrast the well defined 2-fold splitting into $e_g$ and $t_{2g}$ levels is clearly visible for the bulk system ((Fig.~\ref{fig:2}(a) and (b)). Importantly, and this has been pointed out before \cite{HEINRICH_2018,Ruby2016,Saunderson_2022}, splitting the degeneracy of the YSR peaks is fully understood in terms of the crystal-field splitting of the normal state. The key significance in the superconducting state is that as the YSR states form in the spectral gap of the superconductor they sharpen due to minimal hybridizations and their whole spectrum is contained within the energy scale of the superconducting gap. \textcolor{black}{Additionally, the $d_{xy}$ coherence peak is clearly encroaching on the coherence peak of the superconductor. This is a feature of the fact that we are using a multi-orbital description of the full system and has previously been pointed out in investigations involving H-NbSe$_2$.~\cite{Senkpiel_2019}} 

For the adatom, the normal state picture (Fig.~\ref{fig:2}(c)) shows the $d_{xy}$ level to be the most localized, followed by the $d_{x^2-y^2}$, $d_{z^2}$ and finally the degenerate $d_{xz}/d_{yz}$ states. This is a direct result of the differing degrees of hybridization of the different orbitals with the Pb host. The out of plane orbitals hybridize most with the Pb surface atoms below, resulting in a less localized LPDOS. In addition, the $d_{x^2-y^2}$ orbital has a larger overlap with the Pb atoms below as opposed to the $d_{xy}$ orbital due to its orientation with respect to the $x$ and $y$ axis, as depicted in Fig.~\ref{fig:1}(c), again resulting in a reduced localization. In contrast, for the surface atom (Fig. \ref{fig:2}(e)) the $d_{xy}$ states become the least localized, whereas the $d_{xz}/d_{yz}$ orbitals the most. As the substitutional surface impurity replaces a surface Pb atom, the $d_{xy}$ orbital has the largest overlap with the surrounding host, as depicted in Fig.~\ref{fig:1}(f) leading to a broadening of the impurity level.

The change in the surrounding symmetry has a direct impact on the observed YSR states as the relative localization of these states has a direct impact on the relative magnitude of their respective LPDOS at the Fermi energy in the normal state, as is clearly visible in Fig.~\ref{fig:2}. As a result, this has a direct impact on the peak heights of the YSR states in the superconducting state \textcolor{black}{which are non-trivially coupled to the underlying electronic structure due to their multi-orbital nature~\cite{Senkpiel_2019}}.
For the energetic position of the YSR states, theory \cite{Choi_2017,Yu_1965, shiba_1968, Rusinov_1969_superconductivity, HEINRICH_2018, balatsky_2006} predicts that the magnetic moment plays a key  role in the formation of bound states as it affects the exchange splitting which is ultimately responsible for the formation of the YSR states. The magnetic moment for Vanadium in the adatom and surface atom positioning is $3.86 \ \mu_B$ and $3.62 \ \mu_B$ respectively. It is considerably reduced in the case of the surface impurity due to the proximity to Pb atoms. For the adatom (Fig.~\ref{fig:2}(d)) we find a clear four-fold splitting of the states, which is far less pronounced for the surface atom (Fig.~\ref{fig:2}(f)), where we find a splitting mainly into two groups. It appears that the hybridization of the $d_{xy}$ orbitals pointing along the nearest neighbour bonds is the most pronounced and dominates all other effects. This is not only visible for the superconducting state but is equally evident for the normal state (Fig.~\ref{fig:2}(e)). Such a discussion of the symmetry and the local structure will only provide qualitative interpretation, while the exact form or energetic positioning remains hard to predict using qualitative models \cite{HEINRICH_2018}\textcolor{black}{, with some peaks even being observed in the coherence peak of the superconductor. \cite{Senkpiel_2019} } For that reason full material specific calculations are necessary to understand the full picture.

\begin{figure}[t!]
    \centering
    \includegraphics[width=8cm]{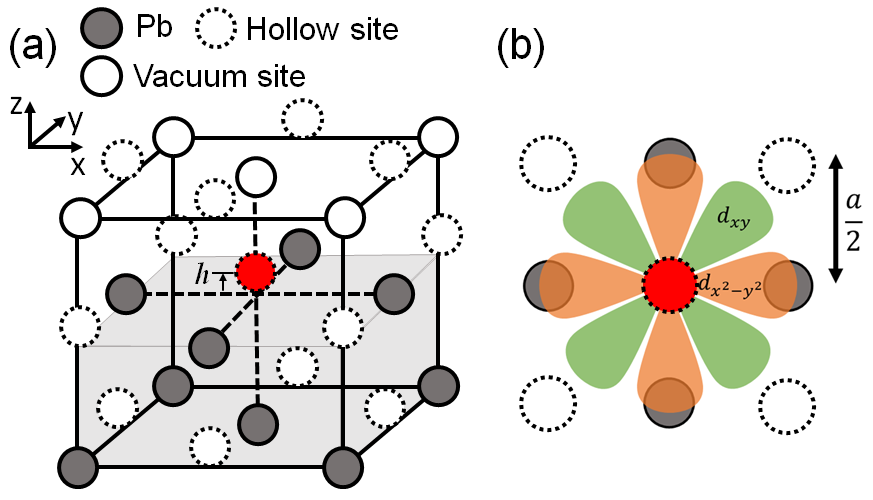}
    \caption{(a) Shows the atomic sites around the impurity atom in units of the atomic spacing, $a = 4.95$~\AA. Grey circles with solid lines represent Pb atom sites, white circles with solid lines represent vacuum sites above the surface, white circles with dashed lines represent hollow sites throughout the Pb and vacuum, and the red disk represents the impurity site. (b) shows the orientation of the d$_{xy}$ (green) and d$_{x^2-y^2}$ (orange) orbitals with respect to the surface layer Pb atoms with hollow site positions marked.}
    \label{fig:4}
\end{figure}

As the true predictive power of any DFT based calculation relies on the analysis of trends we will explore the change in the energetic positioning of the YSR state by varying the impurity atom in the adatom position. In the following, we will consider the full 3$d$ series. 

\subsection{\label{sec:single_imp_comparison} YSR states for the 3$d$ series of impurities}

For V impurities we established some understanding of the energetic ordering and relative peak heights of the YSR states in terms of  the normal state LPDOS around the Fermi energy.

Expanding this analysis to the full 3$d$ series Fig.~\ref{fig:3} shows the development of all YSR states in the adatom position going from Sc to Cu. We left out Zn since it does not develop a magnetic moment (see Table~\ref{tab:mag_moments}) and will be conceptually equivalent to Cu. A few trends are clearly visible. For Sc, Ti, and V the majority spin YSR states are dominant while for Mn, Fe, Co, and Ni the minority spin YSR states are showing the higher peaks. For Cr, the impurity with the largest magnetic moment in the normal state, both spins show almost identical peak heights. This establishes no \textcolor{black}{direct proportionality} between the magnetic moment and the YSR states. \textcolor{black}{It rather shows a nontrivial dependence on the details of the states at the Fermi surface in the normal state. This was similarly found previously~\cite{Saunderson_2022} in bulk calculations calculations. } Furthermore, going from Sc over Ti to V the position of the YSR states move from the left side of the Fermi energy to the right  with the same shift happening in the transition from Mn to Ni for the minority spin YSR states. Again there is no obvious correlation of this trend to the normal state magnetic moment.

\begin{figure*} [ht]
\hspace{-0.7cm}\includegraphics[width=16cm]{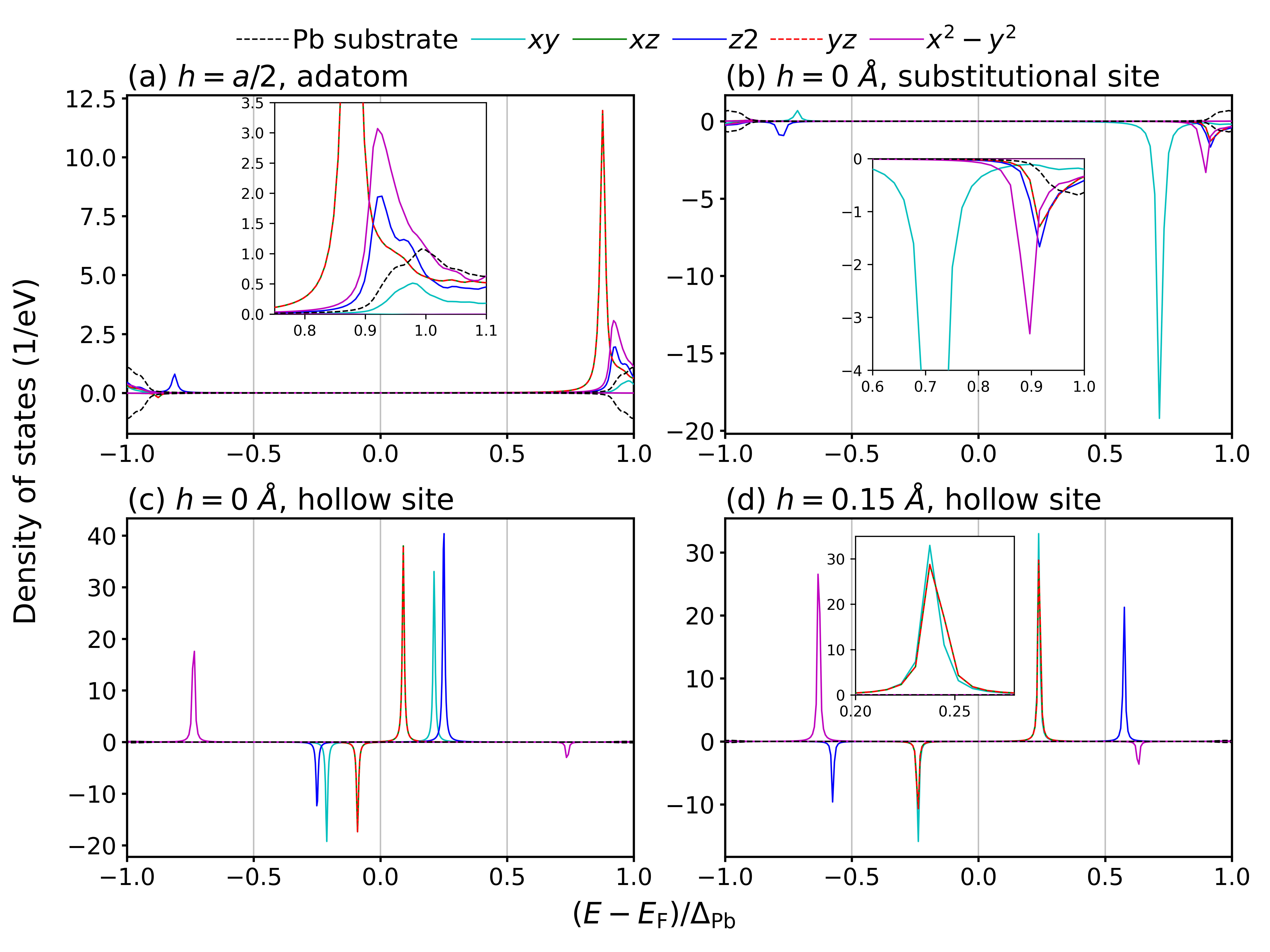}
\caption{Shown is the $d$-orbital resolved LDOS of Manganese as an impurity on a Pb (001) surface exploring the different possible positions, with a) adatom, b) substitutional site c) hollow site, and d) the hollow site position slightly above the surface as found in experiment.}
 \label{fig:5}
\end{figure*}

However, all this can qualitatively be understood in terms of the normal state LDOS of the impurity levels. For Sc to V the majority impurity level moves through the Fermi energy which increases the magnetic moment and shifts the position of the YSR state. For Ti, the system with the largest YSR peak among those three elements, the majority level is closest to the Fermi energy. For Cr, the system with the largest magnetic moment in the normal state, both impurity levels are almost equally far away from the Fermi energy resulting in the small and almost symmetric picture of the YSR states. Moving through from Mn to Ni as the magnetic moment becomes smaller, the minority impurity level moves through the Fermi energy resulting in the shift of the YSR states, again with the highest peak for Co where the impurity level is almost perfectly situated at the Fermi energy. All this highlights the importance of a detailed material specific calculation to understand actual results for the YSR states in real systems. While the specific results can be understood qualitatively from the normal state electronic structure, the subtleties of the electronic structure at the Fermi energy will ultimately determine the precise quantitative results.  

\begin{table}[t]

\begin{ruledtabular}

\begin{tabular}{c| c c c c c c c c c c }

Elements &Sc &Ti & V & Cr & Mn & Fe & Co & Ni & Cu & Zn  \\
\colrule
\rule{0pt}{3ex} Adatom & 0.79 & 2.54 & 3.86 &4.92 &4.81 & 3.50 &2.15 &0.69 & 0 & 0 \\
\rule{0pt}{3ex} Surface & 0 &   2.13&    3.62&    4.71&     4.71&   3.43&   2.03 &   0&  0  &    0\\

\end{tabular}

\end{ruledtabular}

\caption{Magnetic moments (in $\mu_B$) for each element as an adatom or surface atom on a superconducting $(001)$ Pb surface.}

    \label{tab:mag_moments}

\end{table}

In the following we will focus on the special case of Mn impurities, as that was the system considered extensively in experiments.~\cite{Ruby2016}
For Mn impurities the picture changes remarkably in comparison to V (see Fig.~\ref{fig:3}). While the four-fold splitting of the degeneracy of the different orbitals is still visible as was established for the adatom geometry, all states are very close in energy. In addition, all YSR states are energetically very close to the superconducting coherence peaks of the Pb substrate, making it numerically challenging to resolve them at all. Similarly the magnetic moment in itself is not that relevant either given that Mn actually shows a larger moment than the Co adatom impurity (see Table~\ref{tab:mag_moments}). In fact, the results can be understood in terms of the normal state LDOS of the impurity atom. While we discussed at length the local LDOS for the V impurity with the majority impurity level strongly localized at the Fermi level this is not true for the Mn impurity. For Mn the majority level moves further away from the Fermi level, as a direct consequence of the larger magnetic moment. As such the effective difference between the LDOS of minority and majority states at the Fermi energy is less pronounced leading to suppressed YSR states and a reduced difference between the YSR states in the minority and majority channel.

\subsection{\label{sec:single_imp_comparison} Comparison to Experiment}

To effectively explore the YSR states in comparison to the experimental observations~\cite{Ruby2016}, we focus on Pb (001) surfaces with a Mn impurity. However, it turns out that the experimentally observed impurity position is neither the ideal substitutional in-plane position nor the ideal adatom hollow site position discussed so far. The actual position, as indicated by the red impurity site in Fig.~\ref{fig:4}, is the hollow site position but only $0.15$~\AA above the Pb surface.~\cite{Ruby2016}

In order to understand the effect of the crystalline structure on YSR states including the distance between the Mn impurity and the Pb surface, Fig.~\ref{fig:5} shows the decomposition of the LPDOS for the $d$ orbitals for different impurity positions at height $h$, such as $h=a/2$, $0$, and $0.15$~\AA, where $a$ is the lattice constant of Pb. The value of $0.15$~\AA corresponds to the impurity position in the experimental geometry \cite{Ruby2016}. When the Mn atom is at the height of $h=a/2$, all YSR peaks for the different $d$-orbitals strongly hybridize with the superconducting coherence peak of Pb. The $d_{xy}$ orbitals have the highest energy, followed by the $d_{x^2-y^2}$, the $d_{z^2}$ orbitals, and the degenerate $d_{xz}$ and $d_{yz}$ orbitals at the lowest energy. For the impurity at the substitutional site with height $h=0$ (surface atom), the order of energy levels is  $d_{xz/yz}$, $d_{z^2}$, $d_{x^2-y^2}$ and $d_{xy}$ from highest to lowest. At the same height, we additionally calculated the LPDOS for a Mn impurity in the hollow site position (the intersection of the two dashed lines in Fig.~\ref{fig:1}(b)). As the local environment changes dramatically this has a strong effect on all YSR peaks, which move away from the coherence peaks of Pb and split relative to each other. In that case the order of the energy levels is $d_{x^2-y^2}$, $d_{z^2}$, $d_{xy}$, and $d_{xz/yz}$, from highest to lowest. While this order is very close to the experimental finding, as summarized in Table~\ref{tab:exp_comp}, the degenerate $d_{xz}$/$d_{yz}$ and the in-plane $d_{xy}$ still have the wrong relative order and are significantly far apart in energy, which is in contrast to the experimental observation \cite{Ruby2016}.

However, when we move the Mn impurity slightly above the surface as suggested by the experiment~\cite{Ruby2016} with a height of $h=0.15$~\AA the energetic order is in perfect agreement with the experimental observation. In addition to the energetic order even the  narrow gap between $d_{xz,yz}$ and $d_{xy}$ is reproduced in comparison to the experimental observation~\cite{Ruby2016}.

\begin{figure}[t!]
    \centering
    \includegraphics[width=8cm]{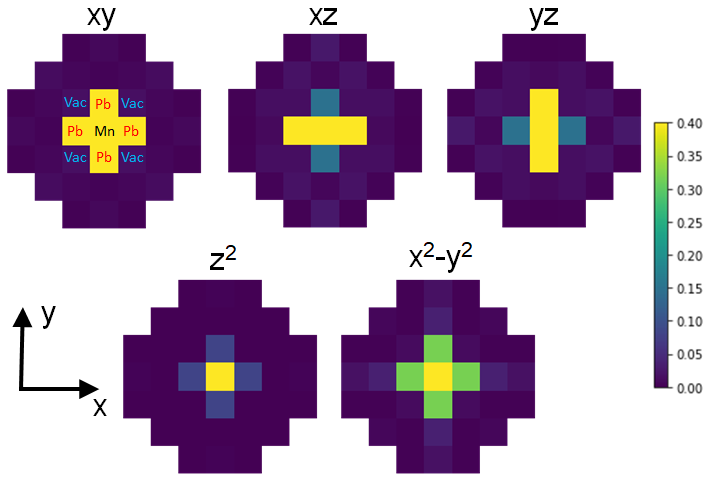}
    \caption{Surface map of the LPDOS induced by an Mn adatom of each peak as shown in figure \ref{fig:4}(d).}
    \label{fig:6}
\end{figure}

In order to make direct contact to the STM experiments by Ruby \textit{et al.} \cite{Ruby2016} we show the atom resolved PDOS in Fig.~\ref{fig:6} for the slightly out of plane hollow site impurity. In each case we fixed the energy to the peak position of the respective orbital shown in Fig.~\ref{fig:6} and visualize the position resolved PDOS. In a first order approximation this is equivalent to the $\frac{dI}{dV}$ spectrum of the STM experiment~\cite{Ruby2016}. The symmetries of each orbital are well reproduced and  comparable to the experimental observation. 

\section{Summary}
 We use our Green's function based BdG method \cite{Saunderson_2020,Saunderson2020b,Saunderson_2022}, complete with collinear magnetism \cite{Saunderson_2022} and substitutional impurities \cite{Saunderson2020b,Saunderson_2022}, to determine how changes to the chemical composition, strength and crystal field around the impurity affect the size and position of the Yu-Shiba-Rusinov states produced with the aim to make contact to experiments such as M. Ruby \textit{et al.} \cite{Ruby2016}. 

Firstly, we compare an impurity embedded in the bulk of superconducting Pb to two possible impurities on the surface of Pb, one being an adatom and the other substituting a Pb atom on the surface. In the bulk case the $d$-orbitals of the impurity split into $e_g$ and $t_{2g}$ states in the normal state and predictably form 2 YSR bound states in the superconducting state. When investigating the impurities on the surface in both cases the $d$-orbitals split into 4 energy levels in the normal state with only $d_{xz}$ and $d_{yz}$ staying degenerate, producing 4 resolvable YSR bound states. The broadening of the orbitally resolved spin polarized peaks in the normal state can be qualitatively explained from determining the level of hybridization the orbitals have with the Pb surface, which, in turn, directly affects the peak height of the subsequent YSR state in the superconducting state.
\begin{table}[t]
\begin{ruledtabular}
\begin{tabular}{cc}
Mn position $h$&
Energy level (high to low)\\
\hline
 & \\
$a/2$  & $d_{xy}\quad d_{x^2-y^2}\quad d_{z^2}\quad d_{yz/yz}$  \\
 & \\
$0$, substitutional site & $ d_{xz/yz}\quad d_{z^2} \quad d_{x^2-y^2}\quad d_{xy} $    \\
 & \\
$0$, hollow site & $d_{x^2-y^2} \quad d_{z^2}\quad d_{xy}\quad d_{xz/yz}$ \\
 & \\
$0.15$~\AA, hollow site & $d_{x^2-y^2}\quad d_{z^2}\quad d_{xz/yz}\quad d_{xy}$ \\
 & \\
Expt. \cite{Ruby2016} & $d_{x^2-y^2}\quad d_{z^2}\quad d_{xz/yz}\quad d_{xy}$  \\
 & \\
\end{tabular}
\end{ruledtabular}
\caption{Energy order of the d orbitals (1-5;low-high), compared with the experimental results of Ruby \textit{et al.} \cite{Ruby2016}.}
    \label{tab:exp_comp}
\end{table}

Secondly, we investigate all 3$d$ transition elements in the adatom position. It is found that the height of the YSR peaks are not only dependent on the broadening of the $d$-orbitals in the normal state, as described in the previous section, but also on the magnitude of the normal density of state $d$-orbitals at the Fermi energy. The position of the YSR states is also addressed. In simplified models it is claimed that the exchange interaction and hence ultimately the size of the magnetic moment of the impurity plays a key role in determining the position of YSR states. We found no \textcolor{black}{direct proportionality between these two quantities, and the relationship is more complex and subtly relies on details of the states at the Fermi level.}

Finally, we investigate the effect of the impurity-surface distance on the energetic position of the YSR peaks. We found a dramatic dependence on the impurity position, where even a change of only $0.15$~\AA~ makes a significant difference. Furthermore, each individual energetic position of the YSR peaks has a unique dependence on the impurity height owing to the different ways in which each orbital hybridizes with the Pb surface. We make direct contact to the experiment for an impurity height above the surface of $0.15$~\AA~ as determined by M. Ruby \textit{et al.} \cite{Ruby2016}. Here, we find an almost perfect agreement with the observed splittings, complete with the real-space density modulations, showing the effectiveness of the first principles calculations in describing the experiment.

To conclude, we model, using a Green's function based technique, how the effect of the chemical composition and position of a magnetic impurity on the surface of a superconductor affects the resulting Yu-Shiba-Rusinov states forming in the spectral gap of the superconductor. Simplified models predict that the energetic position of Yu-Shiba-Rusniov states are directly related to the exchange splitting of the impurity. In this work we show there to be no such simple correlation, instead we find that the exchange splitting only determines the height of the impurity peak which is further influenced by the degree of hybridization of each individual $d$-orbital with the surface. Finally, we make contact to experiments by modelling the exact configuration of a Mn impurity on the surface of Pb as determined experimentally~\cite{Ruby2016}. Here, we find an almost perfect agreement with the experimental results, even obtaining the same minimal energy splitting of the two YSR peaks originating from the $d_{xz/yz}$ and $d_{xy}$ orbitals. We believe this work will provide much-needed clarity in more accurately determining how Yu-Shiba-Rusniov peaks will form, and paves the way for understanding how hybridizing these states in a chain of impurities can combine to form more complex topological order.

\begin{acknowledgments}
This work was carried out using the computational facilities of the Advanced Computing Research Centre, University of Bristol \cite{uob}. M.-H. W. is grateful to support by EPSRC grant EP/N509619/1. The above work was supported by the Centre for Doctoral Training in Condensed Matter Physics, funded by EPSRC Grant No. EP/L015544/1. T.G.S. gratefully acknowledges support by Deutsche Forschungsgemeinschaft (DFG, German Research Foundation) Grant No. TRR 173/2 - 268565370 Spin+X (project A11). M.G. thanks the visiting professorship program of the Centre for Dynamics and Topology at Johannes Gutenberg University Mainz. 
\end{acknowledgments}

\bibliographystyle{apsrev4-2}

%

\end{document}